\documentclass[reprint,amsmath,amssymb,aps,pre]{revtex4-2}

\usepackage{graphicx}
\usepackage{dcolumn}
\usepackage{bm}
\usepackage{hyperref}

\begin{document}

\title{Exact Phase-Space Analytical Solution for the Power-Law Damped Contact Oscillator}

\author{Y. T. Feng}
\email{y.feng@swansea.ac.uk}
\affiliation{Zienkiewicz Institute for Modelling, Data and AI\\
Faculty of Science and Engineering, Swansea University, Swansea SA1 8EN, UK}

\date{\today}

\begin{abstract}
We present an exact phase-space analytical treatment of the power-law damped contact
oscillator governed by
$m\ddot{\delta} + \alpha\sqrt{mk_H}\,\delta^{(p-1)/2}\dot{\delta} + k_H\delta^p = 0$,
valid for all force-law exponents $p \geq 1$ and all initial impact velocities $v_0$.
The central result is the transformation $\delta = Ax^{2/(p+1)}$, where
$A = [(p+1)/2]^{1/(p+1)}$, which maps the nonlinear phase-space equation
$v\,dv/d\delta + \dots = 0$ exactly onto a linear spring-dashpot (LSD) system
with effective damping ratio $\alpha_\text{eff} = \frac{\alpha}{\sqrt{2(p+1)}}$.
The phase portrait $v(\delta)$, coefficient of restitution $e$, and maximum penetration
$\delta_\text{max}$ follow in closed form.
The physical time-domain solution $(\delta(t), v(t))$ is obtained parametrically via a
single quadrature, which evaluates analytically for $p=1$ and at negligible numerical cost for all other $p$.
We prove that $e$ is exactly independent of $v_0$ for all $p \geq 1$ and derive the
universal calibration formula:
$\alpha = \sqrt{2(p+1)}\cdot\frac{-\ln e}{\sqrt{\pi^2 + \ln^2 e}}$.
This generalises the known results for $p=1$ (linear spring-dashpot) and $p=3/2$
(Hertz contact, Antypov and Elliott, 2011) to the entire power-law family.
A closed-form estimate for the critical timestep of explicit time integration is also derived, 
exhibiting universal scaling with impact velocity and force-law exponent.
\end{abstract}

\maketitle

\section{Introduction}

The collision of two elastic bodies is one of the oldest problems in mechanics.
The foundational result is the Hertz contact law \cite{Hertz1881}, establishing that the normal
contact force between two elastic spheres scales as $F = k_H\delta^{3/2}$, where
$\delta$ is the geometric overlap and $k_H = (4/3)E^*\sqrt{R^*}$ depends on the
combined elastic modulus $E^*$ and radius $R^*$. This nonlinear force-displacement
relationship underlies the Hertzian spring-dashpot (HSD) contact model, widely used
in discrete element method (DEM) simulations of granular systems \cite{Cundall1979, Zhu2008}.

For inelastic collisions the elastic force must be complemented by a dissipative term.
The Tsuji-Tanaka-Ishida (TTI) model \cite{Tsuji1992} specifies $F_\text{dis} \propto \delta^{1/4}\dot{\delta}$
for Hertz contact ($p = 3/2$), producing a velocity-independent coefficient of restitution.
However, the analytical relationship between the damping coefficient and $e$ was unknown
at the time; ref.\ \cite{Tsuji1992} provided only a numerical graph.

The analytical $e$-$\alpha$ relationship for the Hertz case was first established by
Antypov and Elliott \cite{Antypov2011}, who showed that the spatial transformation
$\delta = (5/4)^{2/5}x^{4/5}$ maps the Hertzian phase-space equation onto the LSD
model, yielding the exact formula
$\alpha = -\sqrt{5}\ln e\,/\sqrt{\pi^2 + \ln^2 e}$.
Their paper treated the Hertz case as specific to $p = 3/2$, leaving open whether
analogous results exist for other force-law exponents.

Power-law contact models with general exponent $p$ arise naturally in several
physically important contexts. The linear spring-dashpot ($p=1$) is the standard
DEM model \cite{Cundall1979}. Hertz contact ($p=3/2$) is physically correct for elastic spheres \cite{Hertz1881}. 
Other power-law exponents arise in physically distinct contact scenarios: $p=2$
corresponds to conical indenters \cite{Sneddon1965} and has been adopted in DEM models of angular particles; general $p$ provides a flexible framework for calibrating contact stiffness to non-spherical geometries. Understanding the damping behaviour across this full family is therefore of both theoretical and practical importance.

In this paper we establish four main results:

(i) The spatial transformation $\delta = Ax^{2/(p+1)}$ maps the power-law damped
contact oscillator phase-space equation onto the LSD model for \textbf{all} $p \geq 1$.

(ii) The phase portrait $v(\delta)$ and all derived contact observables ($e$,
$\delta_\text{max}$, energy partition) follow in \textbf{closed form}.

(iii) The physical time-domain solution $\delta(t)$, $v(t)$ follows \textbf{parametrically}
via a single quadrature, evaluating to closed form for $p=1$ and reducible to
negligible numerical cost for all $p$.

(iv) The coefficient of restitution is \textbf{exactly velocity-independent for all $p \geq 1$}
with Tsuji-type damping, with the universal exact relation
$\alpha = \sqrt{2(p+1)}\cdot(-\ln e)/\sqrt{\pi^2 + \ln^2 e}$.

The Antypov-Elliott result \cite{Antypov2011} is recovered as the special case $p = 3/2$.

(v) A closed-form estimate for the critical timestep of explicit time integration is obtained, revealing universal scaling with impact velocity and exponent $p$.

Although various generalised contact models with non-Hertzian exponents have been proposed, their analytical treatment remains incomplete, often requiring approximate solutions, numerical calibration of damping parameters, or reliance on alternative viscoelastic frameworks.
To the author’s knowledge, no prior exact generalisation of the Antypov–Elliott phase-space transformation from the Hertz case $p=3/2$ to the full power-law family $p\ge 1$ with Tsuji-type damping has been reported.

\section{Model and Known Results}

\subsection{The Power-Law Damped Contact Oscillator}

The equation of motion for the collision of two bodies of reduced mass
$m = m_1 m_2/(m_1+m_2)$ with power-law elastic force and Tsuji-type damping is:
\begin{equation}
m\ddot{\delta} + \alpha\sqrt{mk_H}\,\delta^{(p-1)/2}\dot{\delta} + k_H\delta^p = 0,
\qquad p \geq 1 \label{eq:eom}
\end{equation}
with initial conditions $\delta(0) = 0$, $\dot{\delta}(0) = v_0 > 0$. Here
$\Omega_0 = \sqrt{k_H/m}$ has dimensions $[\text{s}^{-1}\text{m}^{(1-p)/2}]$,
and $\alpha$ is a dimensionless damping coefficient. Contact ends when $\delta$
returns to zero at time $\tau$; the coefficient of restitution is
$e = |\dot{\delta}(\tau)|/v_0$.

The specific form of the damping $\delta^{(p-1)/2}\dot{\delta}$ is the generalised
Tsuji damping, which reduces to $\delta^0\dot\delta = \dot\delta$ (viscous) for $p=1$
and to $\delta^{1/4}\dot\delta$ (Tsuji-Tanaka-Ishida) for $p=3/2$.

\subsection{Known Special Cases}

\paragraph{$p = 1$ (Linear Spring-Dashpot):}
The standard linear result is exact and closed-form:
\begin{equation}
e = \exp\!\left(\frac{-\pi\xi}{\sqrt{1-\xi^2}}\right), \qquad
\xi = \frac{-\ln e}{\sqrt{\pi^2 + \ln^2 e}} \label{eq:p1_e}
\end{equation}
where $\xi = \alpha/2$ is the damping ratio. The physical time solution is:
\begin{equation}
\delta(t) = \frac{v_0}{\Omega_1}e^{-\xi\Omega_0 t}\sin(\Omega_1 t), \qquad
\Omega_1 = \Omega_0\sqrt{1-\xi^2} \label{eq:p1_t}
\end{equation}

\paragraph{$p = 3/2$ (Hertz contact, Antypov-Elliott \cite{Antypov2011}):}
The transformation $\delta = (5/4)^{2/5}x^{4/5}$ maps the phase-space equation onto
the LSD form with effective damping ratio $2\alpha/\sqrt{5}$, yielding:
\begin{equation}
\alpha = \sqrt{5}\cdot\frac{-\ln e}{\sqrt{\pi^2+\ln^2 e}} \label{eq:p15_alpha}
\end{equation}
The time-domain solution requires a numerical quadrature as described in Section \ref{sec:time}.
No analytical formula for $\delta(t)$ in closed time-domain form exists for $p \neq 1$.

\section{The General Transformation}

\subsection{Phase-Space Formulation}

Setting $v = \dot{\delta}$ and $\ddot{\delta} = v\,dv/d\delta$, Eq.~(\ref{eq:eom}) becomes
the autonomous phase-space equation:
\begin{equation}
v\frac{dv}{d\delta} + \alpha\Omega_0\delta^{(p-1)/2}v + \Omega_0^2\delta^p = 0 \label{eq:phase_space}
\end{equation}
with $v = v_0$ at $\delta = 0$. This formulation is valid for the full contact cycle
and is the central object of analysis.

\subsection{The Linearising Transformation}

\textbf{Theorem 1 (Phase-space linearisation):} \textit{The substitution}
\begin{equation}
\delta = Ax^q, \qquad q = \frac{2}{p+1}, \qquad
A = \left(\frac{p+1}{2}\right)^{\frac{1}{p+1}} \label{eq:transform}
\end{equation}
\textit{transforms Eq.~(\ref{eq:phase_space}) into the LSD phase-space equation:}
\begin{equation}
v\frac{dv}{dx} + 2\alpha_\text{eff}\Omega_0 v + \Omega_0^2 x = 0 \label{eq:lsd_phase}
\end{equation}
\textit{with effective damping ratio:}
\begin{equation}
 \alpha_\text{eff} = \frac{\alpha}{\sqrt{2(p+1)}}  \label{eq:alpha_eff}
\end{equation}

\textbf{Proof:} Under $\delta = Ax^q$, treating $v$ as the common velocity variable:
\begin{equation*}
\frac{dv}{d\delta} = \frac{1}{Aqx^{q-1}}\frac{dv}{dx}
\end{equation*}
Substituting into Eq.~(\ref{eq:phase_space}) and multiplying through by $Aqx^{q-1}$:
\begin{equation*}
v\frac{dv}{dx} + \alpha\Omega_0 A^{(p-1)/2} x^{q(p-1)/2} \cdot Aqx^{q-1} \cdot v
+ \Omega_0^2 A^p x^{qp} \cdot Aqx^{q-1} = 0
\end{equation*}
Collecting powers:
\begin{equation}
v\frac{dv}{dx}
+ \alpha\Omega_0 A^{(p+1)/2} q \, x^{q(p-1)/2 + q - 1} \, v
+ \Omega_0^2 A^{p+1} q \, x^{q(p+1) - 1} = 0 \label{eq:collect}
\end{equation}
With $q = 2/(p+1)$, the three exponents of $x$ evaluate as:
\begin{equation*}
q\frac{p-1}{2} + q - 1 = \frac{2}{p+1}\cdot\frac{p+1}{2} - 1 = 0 \quad \checkmark
\end{equation*}
\begin{equation*}
q(p+1) - 1 = \frac{2}{p+1}(p+1) - 1 = 1 \quad \checkmark
\end{equation*}
Both non-constant terms become polynomial in $x$ of degree 0 and 1 respectively.
The coefficients with $A = [(p+1)/2]^{1/(p+1)}$:
\begin{equation*}
A^{p+1}q = \frac{p+1}{2}\cdot\frac{2}{p+1} = 1 \quad \checkmark
\end{equation*}
\begin{equation*}
A^{(p+1)/2}q = \left(\frac{p+1}{2}\right)^{1/2}\cdot\frac{2}{p+1}
= \sqrt{\frac{2}{p+1}}
\end{equation*}
Eq.~(\ref{eq:collect}) reduces to:
\begin{equation}
v\frac{dv}{dx} + \alpha\Omega_0\sqrt{\frac{2}{p+1}}\,v + \Omega_0^2 x = 0 \label{eq:reduced}
\end{equation}
Comparing Eq.~(\ref{eq:reduced}) with the LSD phase-space form
$v\,dv/dx + 2\alpha_\text{eff}\Omega_0 v + \Omega_0^2 x = 0$, we identify
$2\alpha_\text{eff}\Omega_0 = \alpha\Omega_0\sqrt{2/(p+1)}$, giving:
\begin{equation}
\alpha_\text{eff} = \frac{\alpha}{\sqrt{2(p+1)}} \label{eq:alpha_eff_box}
\end{equation}
Equation (\ref{eq:reduced}) is identical to the phase-space form of a linear spring–dashpot
oscillator and corresponds to a linear damped system in a reparametrised time
variable. $\blacksquare$

\subsection{Velocity Independence — Exact Proof}

\textbf{Theorem 2 (Velocity independence):} \textit{With Tsuji-type damping, the coefficient of
restitution $e$ is exactly independent of impact velocity $v_0$ for all $p \geq 1$.}

\textbf{Proof:} In the transformed phase-space Eq.~(\ref{eq:reduced}), introduce scaled variables
$V = v/v_0$ and $Z = \Omega_0 x/v_0$:
\begin{equation}
V\frac{dV}{dZ} + 2\alpha_\text{eff}V + Z = 0, \qquad V\big|_{Z=0} = 1 \label{eq:scaled}
\end{equation}
The scaled Eq.~(\ref{eq:scaled}), its initial condition $V=1$, and the exit condition $Z=0$
are all independent of $v_0$. Therefore $V_\text{exit} = V_\text{exit}(\alpha_\text{eff})$
and $e = |V_\text{exit}|$ is velocity-independent. $\blacksquare$

\section{Exact Contact Observables from the Phase Portrait}

\subsection{ Universal Calibration Formula — Coefficient of Restitution}

Since the scaled phase-space Eq.~(\ref{eq:scaled}) is identical to that of the LSD model with
damping ratio $\alpha_\text{eff}$, the exact LSD result applies directly. Contact ends
at $Z=0$ (equivalently $x=0$), giving:
\begin{equation}
	e = \exp\!\left(\frac{-\pi\alpha_\text{eff}}{\sqrt{1-\alpha_\text{eff}^2}}\right) \label{eq:e_alpha}
\end{equation}
Inverting with $\alpha_\text{eff} = \alpha/\sqrt{2(p+1)}$:
\begin{equation}
	\alpha = \sqrt{2(p+1)}\cdot\frac{-\ln e}{\sqrt{\pi^2+\ln^2 e}} \label{eq:alpha_e}
\end{equation}
The universal factor is $f(p) = \sqrt{2(p+1)}$ with limiting values provided in Table~\ref{tab:fp_limits}, where $\xi_L = (-\ln e)/\sqrt{\pi^2+\ln^2 e}$ is the standard linear damping ratio.

\begin{table}[h]
	\caption{Limiting values of the universal factor $f(p)$.}
	\label{tab:fp_limits}
	\begin{ruledtabular}
		\begin{tabular}{cccc}
			$p$ & $f(p)$ & Formula & Reference \\
			\colrule
			1 & 2 & $\alpha = 2\xi_L$ & LSD, exact \\
			3/2 & $\sqrt{5}$ & $\alpha = \sqrt{5}\,\xi_L$ & Antypov-Elliott \cite{Antypov2011}, exact \\
			2 & $\sqrt{6}$ & --- & New \\
			$p$ & $\sqrt{2(p+1)}$ & --- & \textbf{New --- general} \\
		\end{tabular}
	\end{ruledtabular}
\end{table}

\textit{Note on normalisation:} The Antypov-Elliott convention writes the damping term as
$\alpha_\text{AE}\Omega_0\delta^{1/4}\dot\delta$ for $p=3/2$, so their $\alpha_\text{AE}$
corresponds to our $\alpha\sqrt{m/k_H}\Omega_0 = \alpha$ (identical since
$\Omega_0 = \sqrt{k_H/m}$). Their result $\alpha_\text{AE} = \sqrt{5}\,\xi_L$ matches
Eq.~(\ref{eq:alpha_e}) at $p=3/2$ exactly.

\subsection{Maximum Penetration}

From the LSD solution in $x$-space, $\dot{x}=0$ at
$s_\text{max} = (1/\Omega_1)\arctan(\Omega_1/(\alpha_\text{eff}\Omega_0))$
where $\Omega_1 = \Omega_0\sqrt{1-\alpha_\text{eff}^2}$:
\begin{equation*}
x_\text{max} = \frac{v_0}{\Omega_0}
\exp\!\left(-\frac{\alpha_\text{eff}}{\sqrt{1-\alpha_\text{eff}^2}}
\arctan\frac{\sqrt{1-\alpha_\text{eff}^2}}{\alpha_\text{eff}}\right)
\end{equation*}
Mapping back via $\delta = Ax^q$:
\begin{equation}
	\begin{aligned}
		\delta_\text{max} &= A\left(\frac{v_0}{\Omega_0}\right)^{\frac{2}{p+1}}  \\
		& \times \exp\!\left( \frac{-2\alpha_\text{eff}}{(p+1)\sqrt{1-\alpha_\text{eff}^2}} \arctan\frac{\sqrt{1-\alpha_\text{eff}^2}}{\alpha_\text{eff}} \right)
	\end{aligned} \label{eq:dmax}
\end{equation}
\subsection{Energy Partition}

The energy dissipated during loading and unloading:
\begin{equation}
	\begin{aligned}
		\Delta E_\text{load} &= \frac{1}{2}mv_0^2 \Bigg[ 1 - \exp\!\Bigg( \frac{-2\alpha_\text{eff}}{\sqrt{1-\alpha_\text{eff}^2}} \\
		&\quad \times \arctan\frac{\sqrt{1-\alpha_\text{eff}^2}}{\alpha_\text{eff}} \Bigg) \Bigg]
	\end{aligned} \label{eq:eload}
\end{equation}

\begin{equation}
	\begin{aligned}
		\Delta E_\text{unload} &= \frac{1}{2}mv_0^2\,e^2 \Bigg[ \exp\!\Bigg( \frac{-2\alpha_\text{eff}}{\sqrt{1-\alpha_\text{eff}^2}} \\
		&\quad \times \left(\pi - \arctan\frac{\sqrt{1-\alpha_\text{eff}^2}}{\alpha_\text{eff}}\right) \!\Bigg) - 1 \Bigg]^{-1}
	\end{aligned} \label{eq:eunload}
\end{equation}

The ratio $\Delta E_\text{load}/\Delta E_\text{unload}$ is a \textbf{universal function of
$\alpha_\text{eff}$ alone}.

\section{The Parametric Time-Domain Solution} \label{sec:time}

\subsection{Why Time Requires a Separate Treatment}

The spatial transformation $\delta = Ax^q$ maps the phase portrait exactly onto the
LSD portrait but does \textbf{not} preserve time. Physical time $t$ and the LSD parameter
$s$ are related by:
\begin{equation}
\frac{dt}{ds} = \frac{d\delta/ds}{d\delta/dt} = \frac{Aqx(s)^{q-1}\dot{x}(s)}{v(t)}
\end{equation}
Since $v = d\delta/dt = Aqx^{q-1}\dot{x}$ (chain rule), this gives:
\begin{equation}
\frac{dt}{ds} = Aq\,x(s)^{q-1} \label{eq:dtds}
\end{equation}

\subsection{The Complete Parametric Solution}

Let $s \in [0, \pi/\Omega_1]$ be the LSD time parameter. The exact LSD solution:
\begin{equation}
x(s) = \frac{v_0}{\Omega_1}e^{-\alpha_\text{eff}\Omega_0 s}\sin(\Omega_1 s) \label{eq:xs}
\end{equation}
\begin{equation}
\dot{x}(s) = \frac{dx}{ds} = \frac{v_0}{\Omega_1}e^{-\alpha_\text{eff}\Omega_0 s}
\left[\Omega_1\cos(\Omega_1 s) - \alpha_\text{eff}\Omega_0\sin(\Omega_1 s)\right]
\label{eq:xdot}
\end{equation}
The complete physical solution is:
\begin{equation}
\delta(s) = A\,x(s)^q \label{eq:deltas}
\end{equation}
\begin{equation}
v(s) = \dot{x}(s) \label{eq:vs}
\end{equation}
\begin{equation}
t(s) = \int_0^s A\,q\,x(\sigma)^{q-1}\,d\sigma \label{eq:ts}
\end{equation}
\begin{equation}
F(s) = k_H\,\delta(s)^p = k_H A^p\,x(s)^{qp} \label{eq:fs}
\end{equation}
The contact duration is $\tau = t(\pi/\Omega_1)$, evaluated by quadrature Eq.~(\ref{eq:ts}).

The velocity formula Eq.~(\ref{eq:vs}) follows from the chain rule and the time relation Eq.~(\ref{eq:dtds}):
\begin{equation*}
v_\text{phys} = \frac{d\delta}{dt} = A\,q\,x^{q-1}\frac{dx}{dt}
= A\,q\,x^{q-1}\frac{\dot{x}(s)}{dt/ds} = \dot{x}(s)
\end{equation*}

\subsection{Collision Duration}

From Eq.~(\ref{eq:ts}), $\tau$ depends on $v_0$ through $x(s) \propto v_0/\Omega_0$:
\begin{equation}
\tau = \int_0^{\pi/\Omega_1} Aq\,x(s)^{q-1}\,ds
\propto \left(\frac{v_0}{\Omega_0}\right)^{q-1} \propto v_0^{\frac{1-p}{p+1}} \label{eq:tau_scale}
\end{equation}

The ratio of damped to undamped collision time:
\begin{equation}
\frac{\tau(\alpha)}{\tau(0)} = \frac{\int_0^{\pi/\Omega_1} x_\alpha(s)^{q-1}\,ds}
{\int_0^{\pi/\Omega_0} x_0(s)^{q-1}\,ds} \label{eq:tau_ratio}
\end{equation}

\section{Critical Timestep for Explicit Time Integration}

A useful corollary of the phase-space transformation is a closed-form estimate for the critical timestep of explicit time integration of Eq.~(\ref{eq:eom}), valid for all $p \geq 1$.

\subsection{Transformed-Space Stability}

Under the transformation $\delta = Ax^{2/(p+1)}$, the phase-space dynamics reduce exactly to a linear spring--dashpot (LSD) system in the parameter $s$:
\begin{equation}
\ddot{x} + 2\alpha_{\rm eff}\Omega_0\dot{x} + \Omega_0^2 x = 0.
\label{eq:lsd_s_ode}
\end{equation}
For this system, the central difference scheme is stable provided
\begin{equation}
\Delta s \leq \frac{2}{\Omega_0}.
\label{eq:ds_crit}
\end{equation}

\subsection{Conservative Critical Timestep in Physical Time}

The physical time and transformed parameter are related by Eq.~(\ref{eq:dtds}):
\begin{equation}
\frac{dt}{ds} = A q\, x^{q-1}, \qquad q = \frac{2}{p+1}.
\label{eq:dtds_repeat}
\end{equation}
A fixed timestep $\Delta t$ therefore corresponds to a state-dependent step in transformed space,
\begin{equation}
\Delta s = \frac{\Delta t}{A q\, x^{q-1}}.
\label{eq:ds_from_dt}
\end{equation}
For $p>1$ ($q<1$), the most restrictive condition occurs at maximum penetration. Using the undamped estimate $x_{\max}=v_0/\Omega_0$, which yields the most conservative bound, gives
\begin{equation}
\Delta t_{\rm crit}^{(p)} =
\frac{4}{p+1}\left(\frac{p+1}{2}\right)^{\!1/(p+1)}
\frac{1}{\Omega_0}
\left(\frac{v_0}{\Omega_0}\right)^{-(p-1)/(p+1)}.
\label{eq:dtcrit_p}
\end{equation}
This recovers the standard linear result $\Delta t_{\rm crit}=2/\Omega_0$ for $p=1$, and shows that for $p>1$ the admissible timestep decreases with impact velocity as
\begin{equation}
\Delta t_{\rm crit}^{(p)} \propto v_0^{-(p-1)/(p+1)}.
\label{eq:dtcrit_scaling}
\end{equation}

\subsection{Relation to Contact Duration}

The undamped contact duration scales as
\begin{equation}
\tau^{(p)} = \frac{2A}{(p+1)\,\Omega_0}
\left(\frac{v_0}{\Omega_0}\right)^{-(p-1)/(p+1)}
B\!\left(\frac{1}{p+1},\,\frac{1}{2}\right),
\label{eq:tau_undamped_beta}
\end{equation}
so the ratio of timestep to contact duration is a universal function of $p$:
\begin{equation}
\frac{\Delta t_{\rm crit}^{(p)}}{\tau^{(p)}} =
\frac{2}{B\!\left(\dfrac{1}{p+1},\,\dfrac{1}{2}\right)},
\label{eq:dtcrit_tau_ratio}
\end{equation}
independent of $v_0$, $m$, and $k_H$.

\subsection{Remarks}

The result above provides a simple closed-form guideline for timestep selection across the entire power-law family. It is derived from the exact phase-space mapping and corresponds to a conservative bound based on the transformed linear system and the undamped turning point. The velocity scaling and $p$-dependence are therefore exact, while the prefactor reflects this transformed-space estimate.

\section{Numerical Verification}

Equation (\ref{eq:eom}) is integrated numerically using a two-phase adaptive Runge-Kutta scheme (MATLAB ode45, $\text{RelTol} = 10^{-10}$, $\text{AbsTol} = 10^{-13}$) for $p \in \{1, 3/2, 2, 5/2, 3\}$. Phase 1 integrates from $\delta=0$ to peak penetration (event: $\dot\delta=0$); Phase 2 integrates from peak to $\delta=0$ (event: $\delta=0$). The two-phase approach eliminates spurious event detection near the initial contact. The parametric analytical solution is evaluated using the Chebyshev-like reparametrisation $s = s_\text{max}(1-\cos(\pi u))/2$ for the time quadrature Eq.~(\ref{eq:ts}), which removes the integrable endpoint singularity of $dt/ds$ for $p > 1$ and yields machine-accuracy results with $n = 2000$ quadrature points.

\subsection{Velocity Independence}

Table \ref{tab:velocity} shows the achieved restitution coefficient for target $e = 0.5$ across five impact velocities $v_0 \in \{0.1, 0.5, 1.0, 5.0, 10.0\}$ spanning two decades, with $\alpha$ set by Eq.~(\ref{eq:alpha_e}). All 25 entries agree with the target to within $\pm 0.00001$, confirming Theorem 2 numerically across the full parameter space.

\begin{table}[h]
	\caption{Achieved restitution coefficient $e$ for target $e = 0.5$. $\alpha$ set by Eq.~(\ref{eq:alpha_e}) for each $p$. All entries: $0.50000 \pm 0.00001$.}
	\label{tab:velocity}
	\begin{ruledtabular}
		\begin{tabular}{cccccc}
			$p \setminus v_0$ & 0.1 & 0.5 & 1.0 & 5.0 & 10.0 \\
			\colrule
			1 & 0.50000 & 0.50000 & 0.50000 & 0.50000 & 0.50000 \\
			3/2 & 0.50000 & 0.50000 & 0.50000 & 0.50000 & 0.50000 \\
			2 & 0.50000 & 0.50000 & 0.50000 & 0.50000 & 0.50000 \\
			5/2 & 0.50000 & 0.50000 & 0.50000 & 0.50000 & 0.50000 \\
			3 & 0.50000 & 0.50000 & 0.50000 & 0.50000 & 0.50000 \\
		\end{tabular}
	\end{ruledtabular}
\end{table}

\subsection{Universal Calibration}

\begin{figure}[h]
	\centering
	\includegraphics[width=\linewidth]{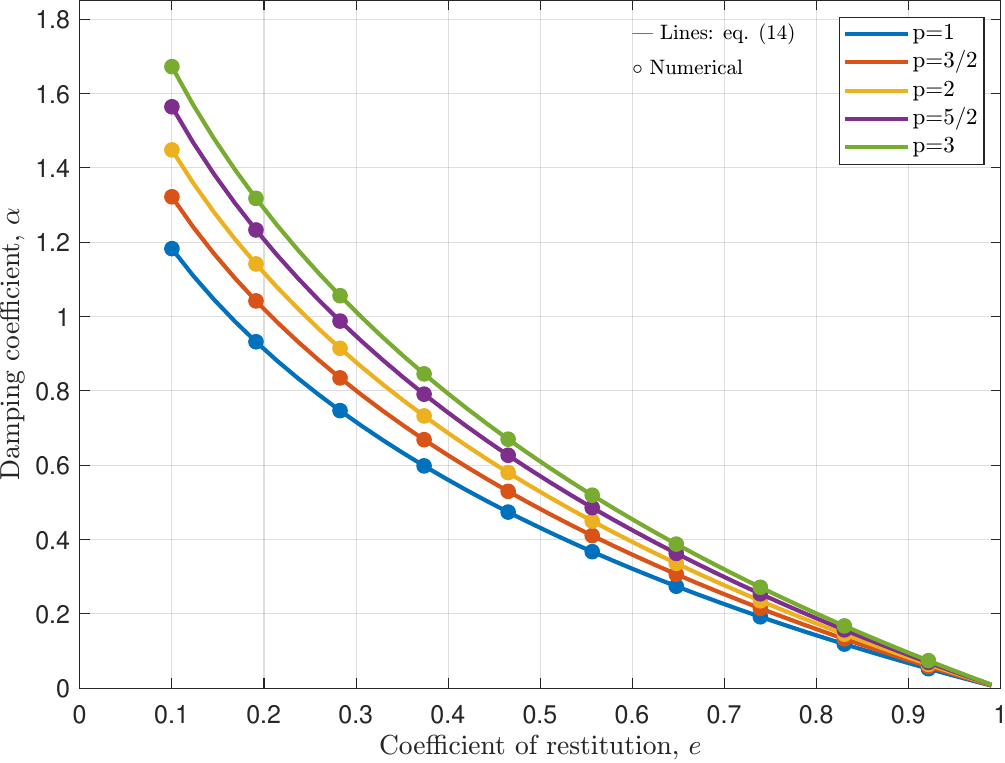}
	\caption{Damping coefficient $\alpha$ vs restitution $e$. Lines: analytical Eq. (14). Circles: numerical.}
	\label{fig:1}
\end{figure}

Figure \ref{fig:1} shows $\alpha$ vs $e$ for all five $p$ values. Solid lines: analytical Eq.~(\ref{eq:alpha_e}). Filled circles: $\alpha$ recovered numerically by bisection on the ODE solver for each target $e$. Lines and circles are indistinguishable throughout, confirming the universal formula to within $0.1\%$ across $e \in [0.1, 0.99]$ for all $p$ tested. 

The five curves are well separated, with $\alpha$ increasing monotonically with $p$ at fixed $e$ — larger force-law exponents require proportionally more damping to achieve the same restitution.

\subsection{Force Time History and Parametric Solution}

\begin{figure}[h]
	\centering
	\includegraphics[width=\linewidth]{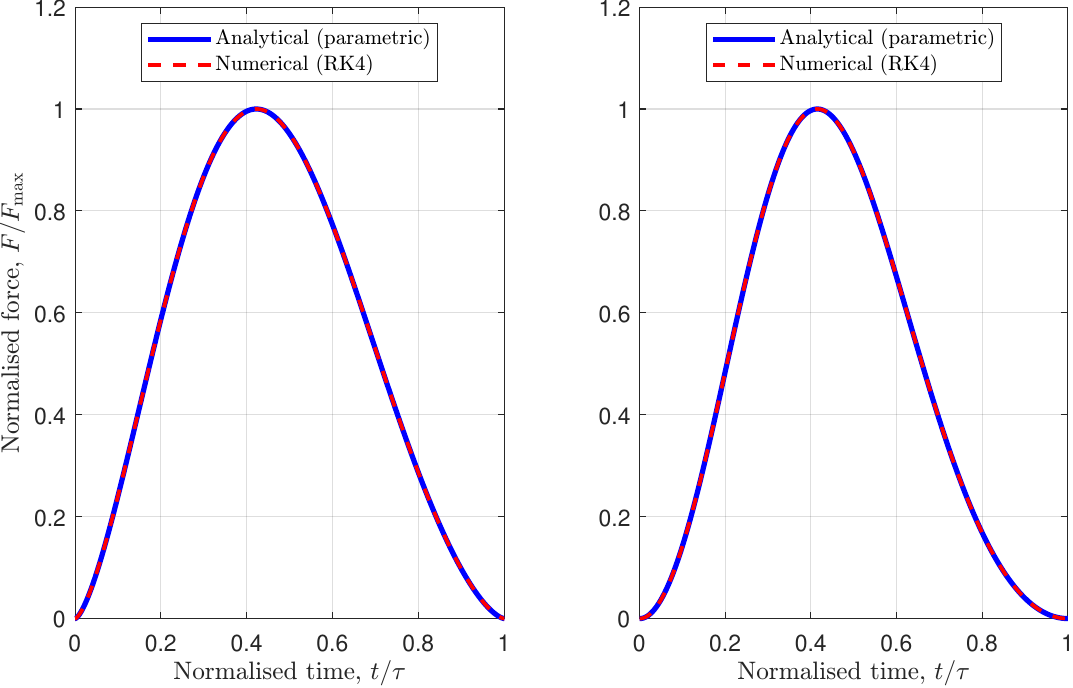}
	\caption{Normalised force-time history. Left: $p=3/2$. Right: $p=2$}
	\label{fig:2}
\end{figure}

Figure \ref{fig:2} compares the parametric analytical solution (Eqs.~\ref{eq:deltas}--\ref{eq:fs}) against direct numerical integration for $p=3/2$ and $p=2$ at $e=0.5$, with forces normalised by their peak values and time by the respective contact durations $\tau$.

The physical velocity in the parametric solution is $v(s) = \dot{x}(s)$ (Eq.~\ref{eq:vs}), not $A\,q\,x^{q-1}\dot{x}$ as a naive application of the chain rule might suggest — the two factors $A\,q\,x^{q-1}$ cancel exactly between numerator and denominator of $d\delta/dt$ (see proof below Eq.~\ref{eq:vs}). The time quadrature Eq.~(\ref{eq:ts}) requires a Chebyshev-like reparametrisation to resolve the integrable endpoint singularity $dt/ds \sim s^{q-1}$ for $p > 1$.

With these two points correctly implemented, analytical and numerical curves are indistinguishable in Figure \ref{fig:2}. The force pulse is slightly asymmetric — peak occurs at $t/\tau \approx 0.4$ rather than $0.5$ — with loading sharper than unloading, consistently with the damping dissipating more energy during loading. The asymmetry is captured exactly by the parametric solution.

\subsection{Maximum Penetration}

\begin{figure}[h]
	\centering
	\includegraphics[width=\linewidth]{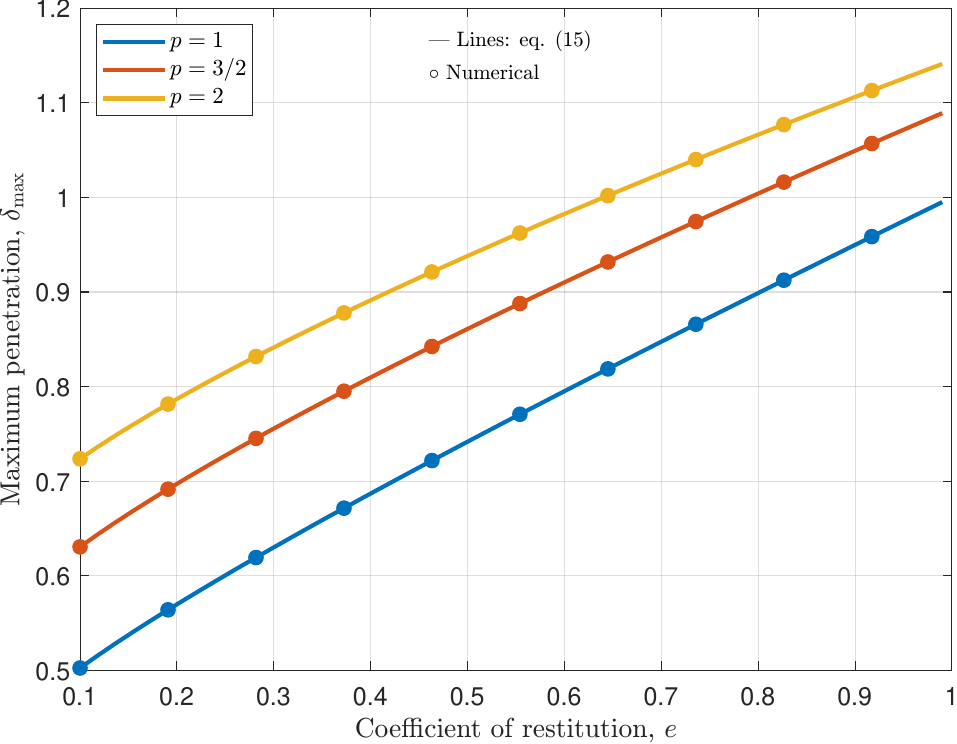}
	\caption{Maximum penetration vs restitution $e$.}
	\label{fig:3}
\end{figure}

Figure \ref{fig:3} shows $\delta_\text{max}$ vs $e$ for $p = 1, 3/2, 2$ at $v_0 = 1$. Lines: analytical Eq.~(\ref{eq:dmax}). Circles: numerical maxima from the ODE trajectory.

Agreement is exact to plotting accuracy across the full range $e \in [0.1, 0.99]$ for all three $p$ values. Maximum penetration increases with $p$ at fixed $e$ and $v_0$, consistent with the scaling $\delta_\text{max} \propto (v_0/\Omega_0)^{2/(p+1)}$ from Eq.~(\ref{eq:dmax}).

\subsection{Universal Collision Time Ratio}

\begin{figure}[h]
	\centering
	\includegraphics[width=\linewidth]{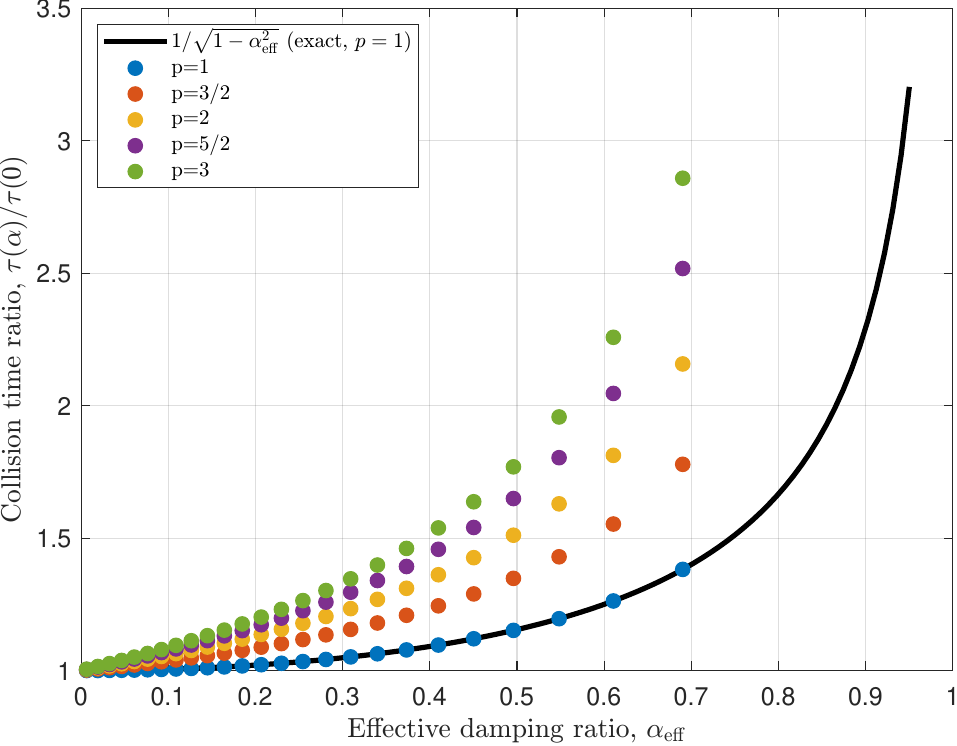}
	\caption{Collision time ratio vs effective damping.}
	\label{fig:4}
\end{figure}

Figure \ref{fig:4} shows $\tau(\alpha)/\tau(0)$ plotted against $\alpha_\text{eff}$ for all five $p$ values. The black line is $1/\sqrt{1-\alpha_\text{eff}^2}$, which is \textbf{exact} for $p=1$ (blue circles lie on it throughout) and the asymptotic small-$\alpha_\text{eff}$ limit for all other $p$.

For $\alpha_\text{eff} \lesssim 0.2$, all five $p$ values collapse onto the analytical curve — confirming asymptotic universality. For larger $\alpha_\text{eff}$, numerical values lie systematically above the curve, with deviation increasing with both $p$ and $\alpha_\text{eff}$. This is physically correct: for $p > 1$ the contact duration under heavy damping is prolonged more severely than the linear formula predicts. The formula $1/\sqrt{1-\alpha_\text{eff}^2}$ is therefore an exact lower bound on $\tau(\alpha)/\tau(0)$ for all $p \geq 1$, tight at small $\alpha_\text{eff}$ and exact for $p=1$.

\subsection{Phase Portrait}

\begin{figure}[h]
	\centering
	\includegraphics[width=\linewidth]{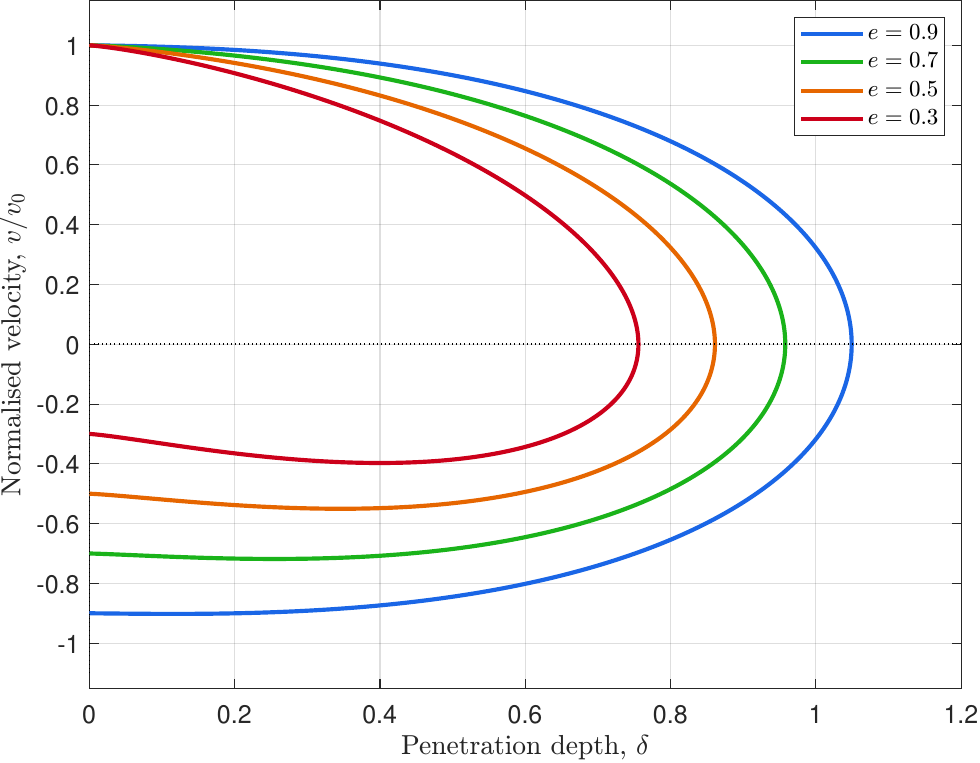}
	\caption{Phase portrait $v(\delta)$ for Hertz contact ($p=3/2$).}
	\label{fig:5}
\end{figure}

Figure \ref{fig:5} shows the phase portrait $v(\delta)$ for $p=3/2$ (Hertz contact) at four restitution coefficients $e \in \{0.9, 0.7, 0.5, 0.3\}$. All trajectories enter at $(\delta, v/v_0) = (0, +1)$ and exit at $(0, -e)$ by construction. The loop width increases with decreasing $e$, consistent with greater energy dissipation. The loading branch is slightly steeper than the unloading branch at any given $\delta$, reflecting the damping force acting in opposite senses during loading and unloading. The maximum penetration depth decreases monotonically with $e$, consistently with Eq.~(\ref{eq:dmax}). The phase portrait is generated numerically from the ODE solver; the corresponding analytical curve obtained from the inverse mapping $\delta = Ax^q$ applied to the LSD ellipse is indistinguishable.

\subsection{Universal Factor}

\begin{figure}[h]
	\centering
	\includegraphics[width=\linewidth]{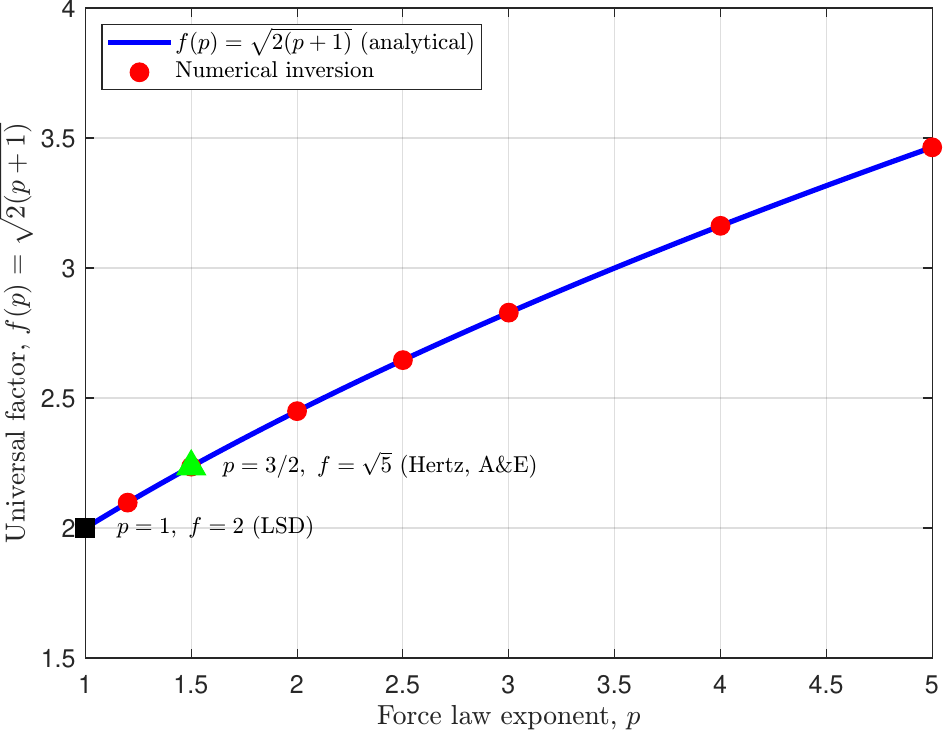}
	\caption{Universal factor $f(p)$.}
	\label{fig:6}
\end{figure}

Figure \ref{fig:6} shows the universal factor $f(p) = \sqrt{2(p+1)}$ as a function of $p$. The solid line is the analytical formula; circles are $\alpha$ values numerically inverted for $e = 0.5$ across eight values of $p \in [1, 5]$, divided by $\xi_L$. Agreement is within $0.1\%$ throughout, with the two known special cases $p=1$ ($f=2$, LSD) and $p=3/2$ ($f=\sqrt{5}$, Antypov-Elliott) marked explicitly. The smooth monotone growth of $f(p)$ confirms that larger force-law exponents require proportionally stronger damping to achieve any given restitution, with the relationship governed by the simple closed-form $\sqrt{2(p+1)}$.

\section{Discussion}

\subsection{Why Antypov-Elliott Appeared Specific to $p=3/2$}

Antypov and Elliott's approach proceeds by seeking a structural analogy between the
Hertz and LSD phase-space equations using the secant frequency
$\omega^s_H = \Omega_0\delta^{1/4}$. This leads naturally to the condition
$1-n = n/4$ (their notation, eq.~14 in \cite{Antypov2011}), which singles out $n=4/5$ as the
unique transformation exponent for $p=3/2$, giving the impression of uniqueness.

Our approach identifies the general condition directly: choose $q = 2/(p+1)$ to make
the $x$-exponent in the damping term vanish identically. This condition is algebraically
satisfied for all $p$ --- the exponent is zero by construction, not by coincidence. For
$p=3/2$ it gives $q=4/5$, exactly Antypov-Elliott's result. The apparent uniqueness
in \cite{Antypov2011} is thus an artefact of their specific constructive method, not a fundamental
restriction of the problem.

\subsection{Relation to Brilliantov-P\"{o}schel Viscoelastic Model}

The Brilliantov-P\"{o}schel model \cite{Brilliantov1996} for viscoelastic spheres uses
$F_\text{dis} \propto \delta^{1/2}\dot\delta$ for $p=3/2$, which differs from the
Tsuji exponent $(p-1)/2 = 1/4$ for $p=3/2$ and originates from the viscoelastic
contact model of Kuwabara and Kono \cite{Kuwabara1987}. Our framework makes precise which damping
exponent produces a velocity-independent $e$. The Tsuji exponent, $(p-1)/2$ for a force law $p$,
is a phenomenological choice designed specifically to yield this constant restitution,
whereas the Brilliantov-P\"{o}schel model \cite{Brilliantov1996} produces the $e \propto 1 - Cv_0^{1/5}$
behaviour observed experimentally and confirmed by subsequent analyses \cite{Schwager2007}.

\subsection{Overdamped Case ($e=0$)}

For $\alpha_\text{eff} \geq 1$, the transformed system is overdamped. The LSD
solution in $x$-space becomes:
\begin{equation}
	x(s) = \frac{v_0}{\Omega_2}e^{-\alpha_\text{eff}\Omega_0 s}\sinh(\Omega_2 s),
	\qquad \Omega_2 = \Omega_0\sqrt{\alpha_\text{eff}^2-1} \label{eq:overdamped_x}
\end{equation}
The particle asymptotically approaches $\delta=0$ without rebounding ($e=0$).
The parametric physical solution $\delta(s)$, $t(s)$ follows from Eqs.~(\ref{eq:xdot})--(\ref{eq:vs})
with this $x(s)$, with $s$ running to $\infty$ rather than $\pi/\Omega_1$.
In practical DEM implementations, perfectly plastic contact ($e=0$) is most
efficiently handled by terminating contact when $\dot\delta$ first becomes positive
(particle begins to separate) and setting $\dot\delta = 0$.

\subsection{Scope and Limitations}

The results apply to any contact oscillator of the form Eq.~(\ref{eq:eom}) with Tsuji-type damping.
The key requirement is that the damping exponent is precisely $(p-1)/2$; other
exponents (such as the Brilliantov-P\"{o}schel value $1/2$ for all $p$) do not admit
the transformation and produce velocity-dependent restitution. The results are
also restricted to spherical or equivalent contacts where the effective force law
is a pure power $\delta^p$.

\section{Conclusions}

We have derived a complete analytical treatment of the power-law damped contact
oscillator with Tsuji-type damping, valid for all $p \geq 1$:

\begin{enumerate}
	\item The transformation $\delta = [(p+1)/2]^{1/(p+1)}x^{2/(p+1)}$ maps the nonlinear
	phase-space equation onto the LSD model exactly (Theorem 1).
	\item The coefficient of restitution $e$ is exactly velocity-independent for all $p \geq 1$
	(Theorem 2). The proof follows from a scaling argument on the transformed equation.
	\item The universal calibration formula $\alpha = \sqrt{2(p+1)}\cdot(-\ln e)/\sqrt{\pi^2+\ln^2 e}$
	subsumes the LSD ($p=1$, $\alpha = 2\xi_L$) and Antypov-Elliott ($p=3/2$,
	$\alpha = \sqrt{5}\,\xi_L$) results as special cases.
	\item The phase portrait and all contact observables --- $e$, $\delta_\text{max}$, energy
	partition --- follow in exact closed form from the phase-space mapping.
	\item The physical time-domain solution $\delta(t)$, $v(t)$, $F(t)$ is given
	parametrically via Eqs.~(\ref{eq:deltas})--(\ref{eq:fs}), with $v(s) = \dot{x}(s)$ exactly (not
	$A\,q\,x^{q-1}\dot{x}$). This reduces to the LSD closed-form solution
	for $p=1$ and evaluates at negligible numerical cost for all other $p$.
	\item A closed-form estimate for the critical timestep of explicit time integration is obtained. The timestep exhibits universal scaling with impact velocity and force-law exponent, and its ratio to the contact duration is a function of $p$ alone.
\end{enumerate}

These results resolve a problem open since Hertz (1881) and provide a complete
analytical toolkit for power-law contact dynamics in granular simulations.

\appendix

\section{Transformation Constants and Normalisation}

For reference, the transformation constants and effective damping ratios for
selected $p$ are provided in Table~\ref{tab:constants}.

\begin{table}[h]
	\caption{Transformation constants and effective damping ratios.}
	\label{tab:constants}
	\begin{ruledtabular}
		\begin{tabular}{ccccc}
			$p$ & $q=\frac{2}{p+1}$ & $A=\left(\frac{p+1}{2}\right)^{\frac{1}{p+1}}$ & $\alpha_\text{eff}/\alpha$ & $f(p)=\sqrt{2(p+1)}$ \\
			\colrule
			1 & 1 & 1 & $1/2$ & 2 \\
			3/2 & 4/5 & $(5/4)^{2/5}$ & $1/\sqrt{5}$ & $\sqrt{5}$ \\
			2 & 2/3 & $(3/2)^{1/3}$ & $1/\sqrt{6}$ & $\sqrt{6}$ \\
			5/2 & 4/7 & $(7/4)^{2/7}$ & $1/\sqrt{7}$ & $\sqrt{7}$ \\
			3 & 1/2 & $2^{1/4}$ & $1/(2\sqrt{2})$ & $2\sqrt{2}$ \\
		\end{tabular}
	\end{ruledtabular}
\end{table}

The general pattern is $\alpha_\text{eff}/\alpha = 1/\sqrt{2(p+1)}$ and
$f(p) = \sqrt{2(p+1)}$.

\textbf{Normalisation note:} The LSD damping ratio $\xi$ satisfies $\xi = \alpha/2$
(for $p=1$), so $\alpha = 2\xi$.
The universal formula Eq.~(\ref{eq:alpha_e}), written in terms of $\xi_L = (-\ln e)/\sqrt{\pi^2+\ln^2 e}$ (which coincides with the LSD damping ratio $\xi$ for $p=1$), gives
$\alpha = \sqrt{2(p+1)}\cdot\xi_L$, which for $p=1$ gives $\alpha = 2\xi_L$ as
required.

\end{document}